\documentclass[12pt, oneside]{amsart}
\pdfoutput=1
\usepackage{geometry}
\usepackage{lmodern}
\allowdisplaybreaks
\usepackage{amsmath}
\usepackage{amssymb}
\usepackage{amsthm}
\usepackage{color}
\usepackage[usenames,dvipsnames]{xcolor}
\usepackage{graphicx}
\usepackage{threeparttable}
\usepackage{subfig}
\usepackage{soul}
\usepackage[countmax]{subfloat}
\usepackage{float}
\usepackage{rotfloat}
\usepackage{comment}
\usepackage{setspace}
\usepackage{esint}
\usepackage{tikz}
\usepackage{multicol}
\usepackage{lscape}
\usepackage{mathtools}
\usepackage[authoryear]{natbib}
\definecolor{MyDarkBlue}{rgb}{0,0.08,0.45}
\usepackage[unicode=true,pdfusetitle, bookmarks=false,bookmarksnumbered=false,bookmarksopen=false, breaklinks=false,pdfborder={0 0 1},colorlinks=true, linkcolor = MyDarkBlue, citecolor = MyDarkBlue]{hyperref}
\usepackage{breakurl}

\PassOptionsToPackage{normalem}{ulem}
\usepackage{ulem}

\makeatletter

\numberwithin{equation}{section}

\newtheorem{theorem}{{\bf\sc Theorem}}

\newtheorem{corollary}{{\bf\sc Corollary}}

\newtheorem{assumption}{{\bf\sc Assumption}}

\newtheorem{lem}{{\bf \sc Lemma}}[section]

\DeclareMathOperator*{\wkto}{\rightsquigarrow}
\providecommand{\E}{\mathrm{E}}

\providecommand{\A}{\mathcal{A}}
\providecommand{\B}{\mathcal{B}}
\providecommand{\var}{\mathrm{var}}

\providecommand{\cov}{\mathrm{cov}}

\providecommand{\Prob}{\mathrm{P}}

\providecommand{\sign}{{\rm sign}}

\renewcommand{\Pr}{\Prob}
\renewcommand{\bar}{\overline}

\renewenvironment{proof}[1][Proof]{\noindent\text{#1.} }{\ \rule{0.5em}{0.5em}}

\providecommand{\barS}{\overline{S}}

\providecommand{\wkto}{\rightsquigarrow}
\providecommand{\Nrm}{\mathcal{N}}

\makeatother
\begin{document}

\title[Affinity Sets]{General Covariance-Based  Conditions for Central Limit Theorems with Dependent Triangular Arrays}

\author{Arun G. Chandrasekhar$^{\ddagger,\S}$}
\author{Matthew O. Jackson$^{\ddagger, \star}$ }
\author{Tyler H. McCormick$^{\diamond,\P}$}
\author{Vydhourie Thiyageswaran$^{\diamond}$}
\date{}

\thanks{$^{\ddagger}$Stanford, Department of Economics}
\thanks{$^{\S}$J-PAL}
\thanks{$^{\star}$Santa Fe Institute}
\thanks{$^{\diamond}$University of Washington, Department of Statistics}
\thanks{$^{\P}$University of Washington, Department of Sociology}

\begin{abstract}
We present a general central limit theorem with simple, easy-to-check covariance-based sufficient conditions for triangular arrays of random vectors when all variables could be interdependent.  The result is constructed from Stein's method, but the conditions are distinct from those of related work. Existing approaches require checking bespoke conditions that in many contexts are difficult to verify scientifically and either (i) impose rigid structure, often difficult to interpret and lacking microfoundations (e.g., strong mixing in random fields) or (ii) allow more flexibility but limit the extent of correlations (e.g., sparsity restrictions in dependency graphs). Our approach, in contrast, permits researchers to work with high-level but intuitive conditions based on overall correlation.  
We show that these covariance conditions nest standard assumptions studied in the literature such as $M$-dependence, mixing random fields, non-mixing autoregressive processes, and dependency graphs, which themselves need not imply each other.  
We apply our result practical settings previously not covered in the literature, such as
 treatment effects with spillovers in more settings than previously admitted, covariance matrices, processes with global dependencies such as epidemic spread and information diffusion, and spatial processes with Mat\'{e}rn dependencies.\\
{\bf Keywords:} Central limit theorem, dependent data, Stein's method
\end{abstract}

\thispagestyle{empty}

\maketitle

\vspace{-20pt}
\section{Introduction}

In many contexts researchers use an interdependent set of random vectors to develop estimators and need to establish whether the estimators are asymptotically normal. In existing results, dependency is modeled in idiosyncratic ways, with perhaps unintuitive if not unappealing conditions to describe assumptions on correlation.   
Further, in many settings\textemdash such as treatment effects with spillovers, epidemic spread, information diffusion, general equilibrium effects, and so on,\textemdash  the correlation is non-zero for any finite set of observations across all random vectors, which is not allowed for in the literature.
 
 We present simple covariance-based sufficient conditions for a central limit theorem to be applied to a triangular array of dependent random vectors.  We use Stein's method \citep{stein1986approximate} to derive three high-level, easy-to-interpret conditions.  Stein's method is widely used in theoretical arguments (in fact, a special case of the argument here first appeared in \citep{chandrasekharj2018}) and our goal is not to advance a new technique for proving limit theorems.  Instead, we present a new approach to proving asymptotic normality that is extremely general, nesting many existing approaches. More importantly, it consists of easily-interpretable conditions that are 
 also easy to check.  Researchers can check our conditions by thinking directly about correlations using calculations that we demonstrate are straightforward in several consequential examples. This feature makes the result accessible to a wider range of researchers without having to derive bespoke limit theorems in every setting.  Our approach also eases restrictions required by existing methods, again doing so in a way that is not idiosyncratically imposed by a specific setting.  
 
 We  
follow 
a literature that uses Stein's method to prove asymptotic normality.  Our goal is not to derive a new proof, but rather to provide much more general conditions that allow for some amount of dependence between all observations, but at the same time are minimal in terms of parametric/shape restrictions, making them flexible and easy to check.  

To review, Stein's method observes that
\[
\E\left[Y f(Y)\right] = \E\left[f'(Y)\right] 
\]
for all continuously differentiable functions if and only if $Y$ has a standard normal distribution. So, when considering normalized sums taking the role of $Y$,  it is enough to show that this equality holds for all such functions asymptotically. \cite{rinott2000normal} and \cite{ross2011fundamentals}, among others, provide a detailed view of the method.
Proving normality using Stein's method typically amounts to checking dependence conditions. \cite{bolthausen1982}, for example, establishes how the probability of a joint set of events differs from independence decays in temporal distance. So, as the mixing coefficients decay fast enough in distance, the proof proceeds by checking that the Stein argument follows under the mixing structure. Distance in time can be generalized to space and, further, to random fields.  Existing literature is organized around a number of such conditions that apply in particular data settings.

A peculiar consequence of this organization, however, is that these conditions generally lack a scientific basis for the assumed dependence structure. 
For example, spatial standard errors are often used\textemdash  e.g., in \cite{froot1989consistent,conley1999gmm,driscoll1998consistent}\textemdash when conducting $Z$-estimation (or GMM). However, in actual applications, for instance agricultural shocks such as rainfall or pests or soil, it is not clear that they follow a specific form of interdependence satisfying $\phi$-mixing with a certain decay rate as invoked in \cite{conley1999gmm} (cross-sectionally) or \cite{driscoll1998consistent} (temporally and cross-sectionally). Surely shocks correlate over space, but it is hard to argue empirically that their correlation matches the required mixing conditions. 

To take a different example, some models of social network formation orient themselves by embedding individuals in a random field to deliver central limit theorems.  Distance in the metric space is, then, inversely proportional to the likelihood of a connection.  Given the structure of the field, researchers can toggle dependence as a function of distance, reducing the size of the covariance sum.  However, the structure of the field has implications for consequential properties of the graph, such as clustering patterns \citep{hoff2002latent, lubold2020identifying}, that then often fail to match the empirical observations.  As in the previous example, the researcher probably has some intuition as to whether graph properties (e.g., clustering) are likely or not, but assessing the reasonableness of specific mixing random field assumptions 
is all but impossible.

As an alternative to embedding variables in a metric space and toggling dependency by distance, a literature on dependency graphs emerged \citep{baldi1989normal,goldstein1996multivariate,ross2011fundamentals} in part to be more flexible on the correlation structure. There, the cost is extreme sparsity: 
 observations have indices in a graph, where those that are not edge-adjacent are independent. This provides a different strategy to apply Stein's method, by creating for each observation a dependency neighborhood. Sufficient sparsity in the graph structure allows a central limit theorem to apply, despite not forcing a time- or space-like structure. Examples include \cite*{ross2011fundamentals,goldstein1996multivariate} and a more general treatment in \cite{chen2004normal}.
 
Both embedding indices in a metric space or using a more unstructured dependency structure are similar in the sense that they constrain the total amount of correlation between the $n$ random variables. In principle there are $\binom{n}{2}$-order components to this sum, but, via mixing conditions or sparsity conditions, this sum is assumed to be of order $n$. However, in many settings---such as treatment effects with spillovers, epidemic spread, information diffusion, or general equilibrium effects---we see dependency that are neither random fields nor are there any conditional independencies. 
The correlation is non-zero for any finite set of observations across all random vectors of interest.  Our approach allows for this general dependence.

In the remainder of this section, we provide a brief overview of our approach, which we formalize in the next section. 
We consider a triangular array of $n$ random vectors $X_{1:n}^n\in \mathbb{R}^p$, which are neither necessarily independent nor identically distributed. We study conditions under which their appropriately normalized sample mean is asymptotically normally disturbed. In principle, at any $n$, all $X_i^n$ and $X_j^n$ can be correlated. The proof follows the well-known Stein's method, though we develop and apply specific bounds for our purpose. To apply Stein's method we first associate each random variable with a set of other random variables with which it could have a higher level of correlation, keeping in mind that it could in principle have some non-zero correlation with all variables. 

We call these \emph{affinity sets}, denoted $\A_{(i,d)}^n$, to capture the other random variables with which $i$ may have high correlations in the   $d$-th dimension. We use the term \emph{affinity set} rather than ``dependency neighborhood'' to emphasize the possibility of high and low non-zero covariance structures with arbitrary groupings that satisfy summation conditions. 
We provide sufficient conditions for asymptotic normality in terms of the total amount of covariance within an affinity set and the total amount of covariance across affinity sets. As long as, in the limit, the amount of interdependence in the overall mean comes from the covariances within affinity sets, asymptotic normality follows. 

This yields substantially weaker conditions than in the previous literature. In \cite{arratia1989two}, the authors present Chen's method \citep{chen1975poisson} for Poisson approximation rather than normality, which has a similar approach to ours in collecting random variables into dependencies. While this results in nice finite sample bounds, these bounds consist of three almost separate pieces, making it less friendly to understanding these bounds together with growing sums of covariance of the samples. In fact, all five of the examples studied in \cite{arratia1989two} are limited to cases where at least one of these pieces is identically zero in the cases where Chen's method succeeds. Many empirically relevant examples do not have such zeros, and so our approach substantially expand the relevant applications. 

To preview our conditions, presented formally below, let us take $p=1$ so $X_i^n$ is scalar, and let $Z_i^n$ be centered, $Z_i^n := X_i^n - \E(X_i^n)$. Let 
\[
\Omega_{n} := \sum_{i = 1}^n \sum_{j\in \A^n_{i} } \cov \left(Z_{i}, Z_{j}\right),
\]
denote the total covariance within the affinity sets. Let also $\mathbf{Z}_{-i}:=\sum_{(j) \notin \A^n_{(i)}} Z_{j}$, the sum of the random variables outside of $i$'s affinity set.
Informally, our conditions are the following.
\begin{enumerate}
\item Within affinity set covariance control:
\[
\sum_i \sum_{j,k\in \mathcal{A}_i^n}\E(Z_j Z_k \cdot |Z_i|) \text{ is small relative to } (\Omega_{n})^{3/2}.
\] The average covariance between two random variables in an affinity set, when weighted by the realized magnitude of the reference variable, and the covariance between the averages given the magnitude of the reference variable, is small relative to the comparably adjusted covariance between the reference variables and their affinity sets.

\item Cross affinity set covariance control: 
\[ \sum_{i,j} \sum_{k\in  \mathcal{A}_i^n,l\in  \mathcal{A}_j^n}\cov(Z_i Z_k, Z_j Z_l) \text{ is small relative to } (\Omega_{n})^{2}.
\]
The average covariance across members of two different affinity sets (weighted by the reference variables) is sufficiently small compared to the squared covariance within affinity sets.

\item Outside affinity set covariance control:
\[ \sum_{i} \E(|\mathbf{Z}_{-i}\E(Z_i\mid \mathbf{Z}_{-i})|) \text{ is small  relative to } \Omega_{n}.
\] 
The average absolute conditional covariance outside of affinity sets is small compared to the covariance within affinity sets.
\end{enumerate}
These conditions are general and easy to check, as we show. They also further simplify in a number of cases such as with positive correlations (as in diffusion models and auto-regressive processes) and with binary variables. 

The advantages of this approach, which we see as most salient for emiprical scientists, are several-fold. First, we do not require a sparse dependency structure at any $n$. That is, there can be non-zero correlation between any pair $X_i,X_j$. Much of the dependency graph literature leverages an independence structure in constructing their bounds and, therefore, the bounds we build are different. 

Second, because of this possibility of non-zero covariance across all random vectors, we organize our bounds through covariance conditions. We are reminded of a discussion in \cite{chen1975poisson} in the context of Poisson approximation. Covariance conditions are easy-to-interpret and check, and from an applied perspective often easier to justify from a microfoundation.

Third, our result is for random vectors, and while the application of the Cram\'er-Wold device is simple in our setting---by the nature of how indexing works---it is useful to have and instructive for a practitioner.

Fourth, our setup nests many of the previous literature's examples, most of which do not nest each other. We illustrate the utility of our central limit theorem through several distinct applications. 
We begin with an example of $M$-dependence in a stochastic process, noting that this implies many other types of mixing, such as $\alpha$-, $\phi$-, or $\rho$- mixing~\citep{Bradley05}.  We then move to random fields, where we show an example with $\alpha$- and $\phi$- dependence.  These mixing approaches require constructing idiosyncratic notions of dependence based on the underlying probability distribution that happen to imply bounds on the covariance function (see, for example,~\citet{Rio93} or~\cite{Rio17}), which, as noted above, is not derived from, and may not match, micro-economic foundations or scientific principles. Our covariance-based arguments are compact and direct, placing restrictions on the covariance explicitly and, thus, in a matter that is salient in the scientific context. They are also general rather than based on a specific model or type of dependence. We also show that our framework is applicable outside the context of mixing, giving examples with non-mixing autoregressive processes, dependency graphs, among other examples.

Fifth, we show that our generalizations permit a wider and more practical set of analyses that were otherwise ruled out or limited in the literature. This includes treatment effects with spillover models, covariance matrices, and things like epidemic and diffusion models. Specifically, we extend the treatment effects with spillovers analysis, as in \cite{aronow2017estimating}, to allow every individual's exposure to treatment to possibly be increasing in every other node's treatment assignment, and nonetheless, the relevant estimator is still asymptotically normally distributed. This case, which is ubiquitous in practice: e.g., in diffusion, epidemic, and financial flow models in principle a shock anywhere could theoretical impact any other node's outcomes (albeit with potentially small odds). But this is assumed away in applied work because conventional central limit theorems do not cover such a case.
We also show how a researcher can model covariance matrices without forcing a random field structure as in \cite{conley1999gmm} or \cite{driscoll1998consistent}. This allows applied researchers to proceed with greater generality, and permits structure across units that do not have a natural ordering such as race, ethnicity, caste, and occupation. 

The next two examples concern diffusion. First, we look at a sub-critical epidemic process with a number of periods longer than the graph's diameter. So, whether an individual is infected is correlated with the infection status of any other individual (assuming a connected, unweighted graph). Again, this practical situation is excluded by the previous central limit theorems in the literature. Second, we look at diffusion in stochastic block models to show that our conditions characterize when asymptotic normality holds and when it does not.

Lastly, we turn to the setting of \cite{zhan2023neural}: the estimation of neural network models with irregular spatial dependence, e.g., Mat\'{e}rn covariance functions. The authors provide the first proof of consistent estimation of the neural network model in this dependent setting. We show that the covariance structure of the residuals, on which the asymptotic distribution of the estimator depends, satisfies our main assumptions and our CLT.

The remainder of the paper is organized as follows, Section~\ref{sec:theorem} proves our main result.  Section~\ref{sec:dependence} shows that the conditions we provide nest several commonly used characterizations of dependence: M-dependence, non-mixing autoregressive processes, random fields, and dependency graphs.  Section~\ref{sec:applications} discusses the process for checking our conditions in several applications (peer effects models, socio-demographic distances, sub-critical diffusion models, stochastic block models, and irregularly observed spatial processes), while also yielding new results.  Section~\ref{sec:discussion} provides a discussion.

\vspace{-20pt}
\section{The Theorem}\label{sec:theorem}

We consider a triangular array of $n$ random variables $X_{1:n}^n \in \mathbb{R}^p$, with entries $X_{i,d}^n$ and $d \in \{1,\ldots,p\}$, each of which has finite variance (possibly varying with $n$).  
We let $Z_i^n\in \mathbb{R}^p$ denote the corresponding de-meaned variables,
\(
Z_{i}^n= X_{i}^n-{\rm E}\left[X_{i}^n\right].
\)
The sum, $S^n \in \mathbb{R}^p$, is given by 
\(
S^{n}:=\sum_{i =1}^n Z_{i}^n.
\) 
We suppress the dependency on $n$ for clarity; writing $X_{i,d}$ unless otherwise needed.

\subsection{Affinity Sets} Each real-valued random variable $X_{i,d}$ has an \emph{affinity set}, denoted $\A_{(i,d)}^n$, which can depend on $n$. We require $(i,d) \in \A_{(i,d)}^n$.  
Heuristically,  $\A^n_{(i,d)}$ includes the indices ${j,d'}$ for which the covariance between  $X_{j,d'}$ and  $X_{i,d}$ is relatively high in magnitude, but not those for which the covariance is low. 

There is no independence requirement at any $n$ and, in fact, our sufficient conditions for the central limit theorem bound the total sums of covariances within and across affinity sets. The precise construction of affinity sets is flexible, as long as these bounds on the respective total sums are respected.

\subsection{The Central Limit Theorem} 
Let $\Omega_n$ be a  $p \times p$ matrix  which houses the bulk of covariance across observations and dimensions, summing across variables all the covariances of each variable and the others in its affinity set:  
\[
\Omega_{n,dd'} := \sum_{i = 1}^n \sum_{(j,d')\in \A^n_{(i,d)} } \cov \left(Z_{i,d}, Z_{j,d'}\right).
\]
This is distinct from a total variance-covariance matrix 
\(
\Sigma_{n,dd'} := \sum_{i=1}^n \sum_{j=1}^n \cov \left(Z_{i,d},Z_{j,d'}\right),
\)
which includes terms outside of $\A^n_{(i,d)}$. 

In what follows, we maintain the assumption that $\left\Vert \Omega_n \right\Vert_F \rightarrow \infty$, where $\left\Vert \cdot \right\Vert_F$ is the Frobenius norm. We also presume that for all $(i,d),$ $\E[|Z_{(i,d)}|^3]/\E[|Z_{(i,d)}|^2]^{3/2}$ is bounded above. Define $\mathbf{Z}_{-i,d}:=\sum_{(j, d') \notin \A^n_{(i,d)}} Z_{j,d'}$, the sum of the random variables outside of $(i,d)$'s affinity set.

Our first assumption is that the the total mass of the variance-covariance is not driven by the covariance between members of a given affinity set neither of which are the reference random variables themselves. That is, given reference variable $X_{i,d}$, the covariance of some $X_{j,d'}$ and $X_{k,d''}$ where both are in the reference variable's affinity set is relatively small in total across all such triples of variables compared to the variance coming from the reference variable and its affinity sets. 
\begin{assumption}[Bound on total weighted-covariance within affinity sets]\label{one}
\[
	\sum_{(i,d); (j,d'),(k,d'') \in \A^n_{(i,d)}} \E  \left[|Z_{i,d}| Z_{j,d'} Z_{k,d''} \right]= o\left(\left( \left\Vert \Omega_n \right\Vert_F  \right)^{3/2}\right).
\]
\end{assumption}

The second assumption is that the total mass of the variance-covariance is not driven by random variables across affinity sets relative to two distinct reference variables. That is, given two random variables $X_{i,d}$ and $X_{j,d'}$, the aggregate amount of weighted covariance between two other random variables---each within one of the reference variables' affinity sets---is small compared to the (squared) variance coming from the reference variable and its affinity sets.
\begin{assumption}[Bound on total weighted-covariance across affinity sets]\label{two}
\[
\sum_{(i,d), (j,d'); (k,d'')\in\A^n_{(i,d)}, (l,\hat{d}) \in\A^n_{(j,d')}} \cov\left(Z_{i,d}Z_{k,d''},Z_{j,d'}Z_{l',\hat d}\right)
	=o\left(\left( \left\Vert \Omega_n \right\Vert_F\right)^{2}\right),
\]
\end{assumption}

The third assumption is that the total mass of variance-covariance is not driven by reference random variables and the variables outside of their affinity sets, again compared to the variance coming from the reference variable and its affinity sets.
\begin{assumption}[Bound on total weighted-covariance from outside of affinity sets]\label{three}
\[
\sum_{(i,d)} \E \left(| \E(Z_{i,d} \mathbf{Z}_{-i,d} \vert \mathbf{Z}_{-i,d})| \right) = \sum_{(i,d)} \E \left(|\mathbf{Z}_{-i,d} \E(Z_{i,d}  \vert \mathbf{Z}_{-i,d})| \right)
	= o\left( \left\Vert \Omega_n \right\Vert_F \right).
\]
\end{assumption}

\bigskip 

These three assumptions imply a central limit theorem.

\begin{theorem}\label{clt}
	If Assumptions \ref{one}-\ref{three} are satisfied, then
	$\Omega_n^{-1/2} S^n \wkto \Nrm(0,I_{p \times p})$.
\end{theorem}

\ 

The proof is provided in the Appendix. The argument follows by applying the Cram\'er-Wold device to the arguments following Stein's method, as~\cite{chandrasekharj2018}  argued for the univariate case. Since the Cram\'er-Wold device requires for all $c \in \mathbb{R}^p$
 fixed in $n$ that the $c$-weighted sum satisfies a central limit theorem \citep{biscio2018note}---that is, $(c'\Omega_nc)^{-1/2}c'S^n \wkto \mathcal{N}(0,1)$---we can consider a problem of $np$ random variables with affinity sets. Then, by checking Assumptions \ref{one}-\ref{three} for the case of $c = 1_{p}$ the result follows.

\ 

An important special case is where the affinity sets are the variables themselves: $\A^n_{(i,d)} = \{(i,d)\}$.  In that case, the conditions simplify to a total bound on the overall sum of covariances across variables (the univariate case is in~\cite{chandrasekharj2018}).  It nests many cases in practice, and we provide an illustration in our second application.

\begin{corollary}
	\label{cor:singleton}
	If  $\A^n_{(i,d)} = \{(i,d)\}$, $\E[Z_{i,d} \mathbf{Z}_{-i,d} \vert \mathbf{Z}_{-i,d}] \geq 0$ for every $(i,d)$, and 
	\begin{itemize}
		\item[(i)] $\sum_{(i,d), (j,d')}  \cov ( Z_{i,d}^2, Z_{j,d'}^2 ) =o\left(\left(\left\Vert \Omega_n \right\Vert_F\right)^{2}\right)$, and
		\item[(ii)]   $\sum_{(i,d) \neq (j,d')}  \cov (Z_{i,d},Z_{j,d'}) = o\left(\left\Vert \Omega_n \right\Vert_F\right)$,
	\end{itemize}
	then $ \Omega_n^{-1/2} S^n \wkto \Nrm(0,I_{p \times p}) $.	
\end{corollary}

If $\E[Z_{i,d} \mathbf{Z}_{-i,d} \vert \mathbf{Z}_{-i,d}] \geq 0$ does not hold,  then (ii) can just be substituted by Assumption \ref{three}.   Also, it is useful to note that, for instance, if $p=1$ and the $X_i$'s are Bernoulli random variables with $\E[X_i]\rightarrow 0$ (uniformly), then condition (ii)  implies condition (i) \citep{chandrasekharj2018}.

\section{Models of Dependence}
\label{sec:dependence}

We first present four applications from the literature that prove asymptotic normality: (i) $M$-dependence, (ii) non-mixing autoregressive processes, (iii) mixing random fields, and (iv) dependency graphs. These examples do not necessarily nest each other, though we do comment on relations between the dependence types in terms of mixing, where relevant.  We can construct affinity sets that meet our conditions in each case. A key distinction in our work is that the conditions we provide are general, rather than specific to a particular model class of dependency type.  We provide a sketch of the core assumptions made in the relevant papers to be self-contained for the reader. We show how these assumptions imply our covariance restrictions and the relative complexity of these setups.

We then present five common applications that are not covered by the previous literature but are covered by our model: (i) peer effects, (ii) covariance estimation with socio-demographic characteristics, (iii) subcritical diffusion processes, (iv) diffusion in stochastic block models, and (v) spatial dependence via a Mat\'{e}rn covariance matrix.

In these examples, we maintain consistent use of notation defined in the previous sections.  The remaining notation, however, is kept consistent only within each subsection.

\subsection{$M$-dependence}

\subsubsection{Environment}
We consider Theorem 2.1 of \cite{romano2000more}. In this application there are real-valued time series data, so $p=1$ (and we drop the index $d$) and $\Omega_n$ is a scalar.  Under Romano and Wolf's setup,  $Z_{n,i}$ and $Z_{n,j}$ are independent if $|i - j| > M$. Here, $\{Z_{n,i}\}$ are mean zero random variables. For convenience of the reader, we include the assumptions made in their paper: Suppose $Z_{n,1}, Z_{n,2}, ..., Z_{n,r}$ is an $M$-dependent sequence of random variables for some $\delta > 0$ and $-1 \leq \gamma < 1$,
\begin{enumerate}
    \item $ \E|Z_{n,i}|^{2 + \delta}  \leq \Delta_n$ for all $i$
    \item $\var \left( \sum_{i=a}^{a+k-1} Z_{n,i} \right)k^{-1-\gamma} \leq K_n$ for all $a$ and $k \geq M$
    \item $\var \left( \sum_{i=1}^{r} Z_{n,i} \right)r^{-1}M^{-\gamma} \geq L_n$
    \item $K_n= O\left(L_n\right)$
    \item $\Delta_n = O\left({L_n}^{1 + \delta/2}\right)$
    \item $M^{1 + (1-\gamma)(1 + 2/\delta)} = o(r)$
\end{enumerate}

\subsubsection{Application of Theorem \ref{clt}} We consider the $M$-ball, $\A_i^n= \{j : |j-i| \leq M \}$. We drop the subscript $n$ in $Z_{n,i}$ for convenience. In this case, $\cov(Z_i, Z_j) = 0$ by independence for all $j$ with $|i - j|>M$, so Assumption \ref{three} is satisfied. Under bounded third and fourth moments, we check the remaining assumptions. Assumption \ref{one} is easily verified:
\begin{align*}
\sum_{i; j, k \in \A_i^n} \E[|Z_i| Z_j Z_k] &= O( M^2 \sum_{i} \E[|Z_i|^3]) = o\left( n^{3/2} M^{3/2} \right) = o\left(\Omega_n^{3/2}\right),
\end{align*} following their Assumption 6.
Our Assumption \ref{two} is satisfied similarly following their  Assumption 6:
\begin{align*}
\sum_{i,j;k\in \A_i^n,l\in \A_j^n} \cov(Z_iZ_k,Z_jZ_l) &= O( \sum_{\mathclap{\substack{i,j: |i - j| \leq M,\\
k: |k - i| \leq M,\\ l: |l - i| \leq 2M}}} \var(Z_i^2))  = O(n \cdot M^3 \cdot \var(Z_i^2)) = o(n^2M^2) =  o(\Omega_n^2).
\end{align*}
This is due to the fact that if they are not within that distance, then automatically it is impossible for the $Z_k$ and $Z_l$ to induce any correlation as well.

Following the hierarchy established in~\cite{bradley2007introduction}, $M$-dependence implies several of these commonly used forms of mixing, such as $\rho$-mixing~\citep{Bradley05}.  It also implies $\phi$-mixing and $\alpha$-mixing in the time series context.  We also given a example using $\alpha$- and $\phi$- mixing in the context of random fields below.  Characterizing mixing in the context of random fields requires imposing restrictions on the dependence between $\sigma$-algebras as the number of points in those sets increases, see~\cite{jenish2009central} or~\cite{Bradley05}.

Note that our conditions generate bounded fourth moment requirements, which is not necessarily a condition invoked in every analysis of $M$-dependent processes in the literature, which sometimes have slightly lower moment requirements. Nonetheless, our results are not intended to provide the tightest bounds, but rather general conditions, spanning various types of dependence, that are easily checkable for most applied settings.

\subsection{Andrews' Non-Mixing Autoregressive Processes}
\subsubsection{Environment}
This application is from \cite{andrews1984non}, which allows interdependence in a time series, but one that does not satisfy strong ($\alpha$-) mixing, in order to clarify the distinction between dependence and mixing. Again, $p=1$.  We take  
\(
X_t= \sum_{l=0}^\infty \rho^l \epsilon_{t-l}, 
\)
where $\rho \in (0,1/2]$.  The $\epsilon_t$'s come from a  Bernoulli distribution with success probability $q$. We define the $Z_t$ as normalized, mean zero, $X_t$. Assume, without loss of generality, that $s > t$, so
\begin{align*}
\cov(Z_t,Z_s) &= \cov(\sum_{l=0}^\infty \rho^l \epsilon_{t-l}, \sum_{k=0}^\infty \rho^k \epsilon_{s-k})
= \rho^M  \sum_{l=0}^\infty \rho^{2l} \var(  \epsilon_{t-l}) = \frac{\rho^M}{1-\rho^2} \cdot q (1-q) 
\end{align*}
where $M = s-t$. For a constant $C$ depending only on $\rho$, we show asymptotic normality of $Z_t$: \( 
\frac{1}{\sqrt{C n q(1-q)}} \sum_t Z_t \wkto \mathcal{N}(0,1). 
\)

\subsubsection{Application of Theorem \ref{clt}}

We begin by verifying our conditions for truncated versions of our random variables and then show that the full result applies. To define what we mean by ``truncation", for each $i \in [n]$, let
\begin{align*}
    X_i^{(D)} = \sum_{j=0}^D \rho^j \epsilon_{i-j}
\end{align*} for some $D.$
Let also $S_n^{(D)} := \sum_{i=1}^n Z_i^{(D)}  \equiv \sum_{i=1}^n (X_i^{(D)} - \mathbb{E}[X_i^{(D)}])$. We then define the affinity sets to be $\mathcal{A}_i = \{k : |i-k| \leq D \}$.

We begin by verifying Assumption \ref{one}:
\begin{align*}
\sum_{i; j, k \in \A_i^n} \E[|Z_i^{(D)}| Z_j^{(D)} Z_k^{(D)}] &= O( D^2 \sum_{i} \E[|Z_i|^3]) = o\left( n^{3/2} D^{3/2} \right) = o\left(\Omega_n^{3/2}\right),
\end{align*}

Next, we verify Assumption \ref{two}:
\begin{eqnarray*}
\sum_{i,j;k\in \A_i^n,l\in \A_j^n} \cov(Z_i^{(D)}Z_k^{(D)},Z_j^{(D)}Z_l^{(D)}) &=& O ( \sum_{\mathclap{\substack{i,j: |i - j| \leq D,\\
k: |k - i| \leq D,\\ l: |l - i| \leq 2D}}} \var({Z_i^{(D)}}^2))  \\
&=& O(n \cdot D^3 \cdot \var({Z_i^{(D)}}^2) = o(n^2D^2) =  o(\Omega_n^2).
\end{eqnarray*}

To verify Assumption \ref{three} is trivial; we have $\sum_i \mathbb{E}[|\mathbf{Z}_{-i}^{(D)} \mathbb{E}[Z^{(D)}_i | \mathbf{Z}_{-i}^{(D)}]| ] = \sum_i \mathbb{E}[|\mathbf{Z}_{-i}^{(D)} \mathbb{E}[Z^{(D)}_i]| ] = 0$. Therefore, we have that the truncated variables ${\left(\Omega_n^{(D)}\right)}^{-1/2}S_n^{(D)} \rightsquigarrow \mathcal{N}(0,1)$, and we would like to show the result in full generality, $\Omega_n^{-1/2}S_n \rightsquigarrow \mathcal{N}(0,1)$. To do that, we begin by writing
\begin{align} \label{decomposition}
    \frac{S_n}{\Omega_n^{1/2}} = \frac{S_n^{(D)}}{{\Omega_n^{(D)}}^{1/2}} \frac{{\Omega_n^{(D)}}^{1/2}}{\Omega_n^{1/2}} + \frac{S_n^{(\bar{D})}}{{\Omega_n}^{1/2}}
\end{align} where $S_n^{(\bar{D})} = \sum_{i}^n (Z_i - Z_i^{(D)}) \equiv \sum_{i}^n Z_i^{(\bar{D})}$.
We will show that $\lim_{D\to \infty} \lim_{n \to \infty} \frac{\Omega_n^{(\bar{D})}}{\Omega_n} = 0$, where $\Omega_n^{(\bar{D})} = \sum_{i,j=1} \cov(Z_i^{(\bar{D})}, Z_j^{(\bar{D})})$, and use this to control the second term in the right-hand-side of the  decomposition given in expression (\ref{decomposition}) above, via Chebyshev's inequality. Then, we will also show that $\lim_{D\to \infty} \lim_{n \to \infty}\frac{{\Omega_n^{(D)}}^{1/2}}{\Omega_n^{1/2}} = 1$, which will result in the weak convergence of the first term on the right-hand-side of (\ref{decomposition}) to $\mathcal{N}(0,1).$
We start with 
\begin{align*}
    \lim_{D\to \infty} \lim_{n \to \infty} \frac{\Omega_n^{(\bar{D})}}{\Omega_n} &= \lim_{D\to \infty} \lim_{n \to \infty} \frac{\sum_{i,j} \cov(Z_i^{(\bar{D})}, Z_j^{(\bar{D})})}{\sum_{i,j} \cov(Z_i, Z_j)} \\
    &= \lim_{D\to \infty} \lim_{n \to \infty} \frac{\sum_{i,j}  \rho^{2D + 2 + | i-j|} \sum_{k=0}^\infty \rho^{2k} q (1-q)}{\sum_{i,j} \rho^{|i-j|} \sum_{k=0}^\infty \rho^{2k} q(1-q)} \\
    &= \lim_{D\to \infty} \lim_{n \to \infty} \frac{\sum_{i,j}  \rho^{2D + 2 + | i-j|}}{\sum_{i,j} \rho^{|i-j|}}= 0.
\end{align*}

The final line comes from observing that from
${\sum_{i,j}  \rho^{2D + 2 + | i-j|}}{\sum_{i,j} \rho^{|i-j|} }$,
the denominator $\sum_{i,j} \rho^{|i-j|} = \Omega(n)$, i.e. there exists some constant $c$ such that $\sum_{i,j} \rho^{|i-j|}  \geq c n$. This is because $\sum_{i,j} \rho^{|i-j|} \geq \sum_{i} (\sum_{k=1}^{n/2} \rho^{k}) = \sum_{i} ((1-\rho^{n/2})/(1-\rho)) \geq n c$ for some constant $c > 0$. Additionally, for any fixed $\epsilon > 0$, there exists a $D>0$ such that the numerator $\sum_{i,j}  \rho^{2D + 2 + | i-j|} = O(\epsilon n).$

Together with Chebyshev's inequality, we have that
\begin{align} \label{tail_term}
    \lim_{D \to \infty } \lim_{n \to \infty}\mathbb{P}\left(\left|\frac{S_n^{(\bar{D})}}{{\Omega_n}^{1/2}}\right| > \epsilon \right) \leq \lim_{D \to \infty } \lim_{n \to \infty} \frac{{\Omega_n^{(\bar{D})}}}{\epsilon^2 \Omega_n} = 0,
\end{align} and this gives us convergence in probability of the second term in \ref{decomposition} to zero.

Next, we show that $\lim_{D\to \infty} \lim_{n \to \infty}\frac{{\Omega_n^{(D)}}^{1/2}}{\Omega_n^{1/2}} = 1$. We begin by considering
\begin{align} \label{ratio}
    \frac{\Omega_n^{(D)}}{\Omega_n} &=  \frac{\sum_{i,j} \cov(Z_i^{(D)}, Z_j^{(D)})}{\sum_{i,j} \cov(Z_i, Z_j)} \notag = \frac{\sum_{i,j} \rho^{|i-j|} \sum_{k=0}^{D - |i-j|} \rho^{2k}q(1-q)}{\sum_{i,j} \rho^{|i-j|} \sum_{k=0}^\infty \rho^{2k} q(1-1)} \notag  \\
    &= 1 - \frac{\sum_{i,j} \rho^{|i-j|} \sum_{k =D-|i-j| + 1 \vee 0}^\infty \rho^{2k}}{\sum_{i,j} \rho^{|i-j|} \sum_{k=0}^\infty \rho^{2k}} \notag \\
    &= 1 - \frac{\sum_{i,j:|i-j| \leq D} \rho^{2D-|i-j| + 2} }{\sum_{i,j} \rho^{|i-j|} } - \frac{\sum_{i,j:|i-j| > D} \rho^{|i-j|} }{\sum_{i,j} \rho^{|i-j|} } 
\end{align}

We now consider the second term in the above expression, (\ref{ratio}), ${\sum_{i,j:|i-j| \leq D} \rho^{2D-|i-j| + 2} }/{\sum_{i,j} \rho^{|i-j|} }$.
The lower bound $\sum_{i,j} \rho^{|i-j|} \sum_{k=0}^\infty \rho^{2k} = \Omega (n)$ (i.e. there exists some constant $c$ such that $\sum_{i,j} \rho^{|i-j|} \geq c n$). For a fixed $\epsilon > 0$, there exists a $D > 0$ such that $\rho^{d} < \epsilon$ for all $d \geq D$. Therefore, we have that $\sum_{i,j} \rho^{|i-j|} \sum_{k =D-|i-j| + 1}^\infty \rho^{2k} = O( n \epsilon)$. Hence, 
\begin{align*}
       \frac{\sum_{i,j:|i-j| \leq D} \rho^{2D-|i-j| + 2} }{\sum_{i,j} \rho^{|i-j|} } = O(\epsilon).
\end{align*}

Next, we consider the third term in \ref{ratio}
\begin{align*}
    \frac{\sum_{i,j:|i-j| > D} \rho^{|i-j|} }{\sum_{i,j} \rho^{|i-j|} } = \frac{\sum_{i,j} \rho^{D+ 1}\rho^{j} }{\sum_{i,j} \rho^{|i-j|} }.
\end{align*}
Once again, we have the lower bound $\sum_{i,j} \rho^{|i-j|} = \Omega (n)$ (i.e. there exists some constant $c$ such that $\sum_{i,j} \rho^{|i-j|} \geq c n$), and for a fixed $\epsilon > 0$, there exists a $D > 0$ such that $\rho^{d} < \epsilon$ for all $d \geq D$. Therefore, we have that $\sum_{i,j} \rho^{D+ 1}\rho^{j} = O( n \epsilon)$, and hence, 
\begin{align*}
        \frac{\sum_{i,j:|i-j| > D} \rho^{|i-j|} }{\sum_{i,j} \rho^{|i-j|} } = O(\epsilon).
\end{align*}

Putting all of this together gives us $\lim_{D \to \infty} \lim_{n \to \infty} \frac{{\Omega_n^{(D)}}^{1/2}}{{\Omega_n}^{1/2}} = 1$. 
Finally, we have that the first term in \ref{decomposition}, $\frac{S_n^{(D)}}{{\Omega_n^{(D)}}^{1/2}} \frac{{\Omega_n^{(D)}}^{1/2}}{\Omega_n^{1/2}} \rightsquigarrow N(0,1)$ since $\lim_{D \to \infty} \lim_{n \to \infty}\frac{{\Omega_n^{(D)}}^{1/2}}{\Omega_n^{1/2}} = 1$. Therefore, together with \ref{tail_term}, we have that $\frac{S_n}{\Omega_n^{1/2}} \rightsquigarrow \mathcal{N}(0,1)$, and the proof is complete.

\subsection{Random Fields}\label{rf}

\subsubsection{Environment}
This example nests many time series and spatial mixing models. Take the setting of  \cite{jenish2009central},  Theorem 1. Their setting has either $\phi$- or $\alpha$- mixing in random fields, allowing for non-stationarity and asymptotically unbounded second moments. They treat real mean-zero random field arrays $\{ Z_{i,n}; i \in D_n \subseteq \mathbb{R}^d,  n \in \mathbb{N}\}$, where each pair of elements $i,j$ has some minimum distance $\rho(i,j) \geq \rho_0 > 0$, where $\rho(i,j) := \max_{1 \leq l \leq d} | i_l - j_l |$, between them. At each point on the lattice, there is a real-valued random variable drawn, so $p = 1$. The authors assume (see their Assumptions 2 and 5) a version of uniform integrability that allows for asymptotically unbounded second moments, while maintaining that no single variance summand dominates by scaling $X_{i,n} := Z_{i,n}/{\max_{i \in D_n} c_{i,n}}$ so that $X_{i,n}$ is uniformly integrable in $L_2$. They also assume (Assumption 3, restated below) conditions on the inverse function $\alpha_{inv}$ on mixing coefficients $\alpha$ (their Assumption 3) and $\phi$ (their Assumption 4) together with the tail quantile functions $Q_{i,n}$ (where $Q_X(u) := \inf \{x : F_X(x) \geq 1 - u \}$ where $F_X$ is the cumulative distribution function for the random variable $X$), requiring nice trade-off conditions between the two, such that under $\alpha$-mixing decaying at a rate $O(\rho^{d+\delta})$ for some $\delta >0$, $\sum_{m=1}^\infty m^{d-1} \sup_n \alpha_{k,l,n} (\rho) < \infty$ for all $k + l \leq 4$, and $\sup_n \sup_{i \in D_n} \int_0^1 \alpha^d_{inv}(u) Q_{i,n}(u)du$ tends to zero in the limit of upper quantiles. Restating the assumptions made in their paper: 
\begin{itemize}
    \item Assumption 2: $ \lim_{k \to \infty} \sup_n \sup_{i \in D_n} \E[|Z_{i,n}/c_{i,n}|^{2} \mathbf{1} \{|Z_{i,n}/c_{i,n}| > k \}] =0$ for $c_{i,n} \in \mathbb{R}^+$
    \item Assumption 3: The following conditions must be satisfied by the $\alpha$-mixing coefficients:
    \begin{enumerate}
        \item $\lim_{k \to \infty} \sup_n \sup_{i \in D_n} \int_0^1 \alpha_{inv}^d(u) \left(Q_{|Z_{i,n}/c_{i,n}|\mathbf{1}\{Z_{i,n}/c_{i,n} > k\}} \right)^2 du = 0$
        \item $\sum_{m=1}^\infty m^{d-1} \sup_{n} \alpha_{k,l,n}(m) < \infty$ for $k+l \leq 4$ where 
        $$\alpha_{k,l,n}(r) = \sup (\alpha_n(U,V), |U| \leq k, |V| \leq l, \rho(U,V) \geq r) $$
        
        \item $\sup_{n} \alpha_{1,\infty,n}(m) = O\left( m^{-d - \delta}\right)$ 
    \end{enumerate}
    \item Assumption 4: The following conditions must be satisfied by the $\phi$-mixing coefficients:
    \begin{enumerate}
        \item $\sum_{m=1}^\infty m^{d-1}\bar{\phi}_{1,1}^{1/2}(m)<\infty$
        \item $\sum_{m=1}^\infty m^{d-1}\bar{\phi}_{k,l}(m)<\infty$ for $k+1\leq 4$
        \item $\bar{\phi}_{1,\infty}(m)=\mathcal{O}(m^{-d-\epsilon})$ for some $\epsilon>0$
    \end{enumerate}
    \item Assumption 5: $\lim \inf_{n \to \infty} |D_n|^{-1} {M_n}^{-2} \sigma_n^2 > 0$
\end{itemize}
\subsubsection{Application of Theorem \ref{clt}} In the following, we assume that the $Z_{i,n}$s have bounded second moments (otherwise, we can replace them with their scaled versions (see above), and the results should go through under bounded third and fourth moments). Here, for any $\epsilon > 0$, we take $\A_i^n= \{j : \rho(i,j)  \leq K^i(\epsilon_n) \}$, where $K^i$ is a non-increasing function. That is, we pick $K^i(\epsilon_n)$ to be large enough, and this can be decided by understanding the cumulative distribution function of the random variables. %

By Assumption 3 (and Proposition B.10), and Lemma B.1 (a) in \cite{jenish2009central}, we know that for any $k \neq i$ such that $\rho(i,k) \geq K^i(\epsilon_n)$, we have that for some constant $C$, 
\begin{align} \label{eps_req_rf}
    C \int_0^{\bar{\alpha}_{1,1}(\rho(i,k))} Q_{i,n}^2(u) du \leq \epsilon_n.
\end{align}
In particular, we note that we can pick $K^i(\epsilon_n)$ by observing that $\epsilon_n$ above satisfies $O(1/\rho(i,k)^r)$ for $r > 1.$

Taking $\epsilon_n = \omega( \frac{1}{\rho_0^d n^{\gamma}})$ allows control of the size of the affinity sets. Indeed, via a packing number calculation, we see that while this allows $K^i(\epsilon_n)$ to grow with $n$, it grows more slowly than $n$. Specifically, taking $\epsilon_n = \omega\left( \frac{1}{\rho_0^dn^\gamma}\right)$ for any $0 < \gamma < 1$, and since $\delta > 0$ together with their Assumption 3, we have,
\begin{align*}
    \left(\frac{K^i(\epsilon_n)}{\rho_0}\right)^d = \left(\frac{(1/\epsilon_n)^{-(d + \delta)}}{\rho_0}\right)^d < \left(\frac{1}{\epsilon_n \rho_0^d} \right) = o(n)
\end{align*}

Now, we verify that our key conditions are satisfied in this setting. We write $K:= \max_i K^i(\epsilon_n)$, and first, we check Assumption \ref{one}:
\begin{align*}
    \sum_{i; j, k \in \A_i^n} \E[|Z_{i,n}| Z_{j,n} Z_{k,n}] &= O(K^2 \sum_{i} \E[|Z_{i,n}|^3]) = O(K^2n) = o(n^{3/2}K^{3/2})= o\left(\Omega_n^{3/2}\right)
\end{align*} since $\Omega_n^{3/2} = (\sum_{i} \sum_{j \in \A_i^n} \E[Z_iZ_j])^{3/2}.$ The second inequality holds by the algorithm-geometric mean inequality.  The remaining argument relies on rearranging the summations and using the growth rate of $K$, i.e. $K = o(n)$ in the third equality, as defined above.

Next, we check that Assumption \ref{two} is satisfied using similar arguments and assuming bounded fourth moments: 
\begin{align*}
\sum_{i,j;k\in \A_i^n,l\in \A_j^n} \cov(Z_{i,n}Z_{k,n},Z_{j,n}Z_{l,n}) &= O( \sum_{\mathclap{\substack{i,j: |i - j| \leq K,\\
k: |k - i| \leq K,\\
l: |l - i| \leq 2K}}} \E(Z_{i,n}^4))  
= O(n \cdot K^3 \cdot \E(Z_{i,n}^4)) =  o(\Omega_n^2)
\end{align*}
where the first equality comes from the covariance terms within the affinity sets dominating those with outside the affinity sets (see the following verification of Assumption \ref{three}).

Finally, we check Assumption \ref{three}. 
Together with \ref{eps_req_rf}, taking $\epsilon_n = \omega( \frac{1}{\rho_0^dn^\gamma})$ where $\gamma = 1 - \beta$ for arbitrarily small $\beta > 0$, we have, $\frac{n}{K^{1 + 1/(d+\delta)}} = o(1)$. Thus, defining the random variable $\xi_{i,-i}$ such that $\xi_{i,-i}=1$ if $\E(Z_{i,n} | \mathbf{Z}_{-i,n}) > 0$, and $\xi_{i,-i}=-1$ otherwise just as in \cite{rio2013inequalities}, we have 
\begin{align*}
    \sum_{i} \E ( \left|\E (Z_{i,n} \mathbf{Z}_{-i,n} | \mathbf{Z}_{-i,n})\right|  ) &= \sum_{i} \sum_{j \notin \mathcal{A}_i} \cov(\xi_{i,-i} Z_{j,n}, Z_{i,n}) \\
    &\leq 2 \sum_{i} \sum_{j \notin \mathcal{A}_i} \int_0^{\bar{\alpha}(\rho(i,j))} Q_{i,n}(u) Q_{j,n}(u) du \\
    &= O(n(n-K) \epsilon_n) = O(n(n-K) K^{-1/(2 + \delta)}) = o(nK)=
 o\left(\Omega_n\right),
\end{align*} by Rio's inequality \cite{Rio93}.

\subsection{Dependency Graphs and \cite{chen2004normal}} Next, we consider dependency graphs. There is an undirected, unweighted graph $G$ with dependency neighborhoods $N_i := \{j : \ G_{ij}=1\}$ such that $Z_i$ is independent of all $Z_j$ for $j\notin N_i$  \citep{baldi1989normal,chen2004normal,ross2011fundamentals}. Let $\A_i^n =\{j: \ G_{ij}=1\}$.  Denote the maximum cardinality of these to be $D_n$. In \cite{ross2011fundamentals} (see Theorem 3.6), together with a bounded fourth moment assumption, we see that the conditions there imply the conditions here. Indeed, we see that
\begin{align*}
   \sum_{i; j,k \in \A_i^n} \E  \left[|Z_i| Z_{j} Z_{k} \right] &\leq D^2 \sum_{i=1}^n \E  \left[Z_i^3 \right] ,
\end{align*}
and for $\Omega_n^{-1/2}S_n \wkto N(0,1)$, we need $D^2 \sum_{i=1}^n \E  \left[Z_i^3 \right] \leq  o\left(\Omega_n^{3/2}\right)$ in \cite{ross2011fundamentals} (Theorem 3.6), and hence Assumption \ref{one} is satisfied.
Similarly, 
\begin{align*}
    \sum_{\mathclap{i,i'; j\in\A\left(i,n\right),j'\in\A_i^n}} \cov\left(Z_{i}Z_{j},Z_{i'}Z_{j'}\right) = O(D^3 \sum_{i=1}^n \E [Z_i^4]),
\end{align*} and for $\Omega_n^{-1/2}S_n \wkto N(0,1)$, we need $D^3 \sum_{i=1}^n \E [Z_i^4] \leq o\left(\left(\Omega_n\right)^{2}\right)$ in \cite{ross2011fundamentals} (Theorem 3.6), so Assumption \ref{two} is satisfied. Assumption \ref{three} holds by definition of the dependency neighborhoods.

Now, we consider \cite{chen2004normal}. In particular, we consider their weakest assumption, LD1: Given index set $\mathcal{I}$, for any $i \in \mathcal{I}$, there exists an $A_i \in \mathcal{I}$ such that $X_i$ and $X_{A_i^C}$ are independent. The affinity sets can be defined by the complement of the independence sets. So  $\A^n_i = \{j: Z_j \text{ is not independent of } Z_i\}$, which is similar to the dependency graphs setting. The goals of their paper are different. They develop finite-sample Berry-Esseen bounds, with bounded $p$-th moments, where $2 < p \leq 3$. This is different from our approach. In our paper, we focus on covariance conditions in the asymptotics, and collect relatively more dependent sets along the triangular array.

\section{Applications}
\label{sec:applications}
\subsection{Peer Effects Models}
\subsubsection{Environment} We now turn to an example of treatment effects with spillovers.  
Consider
 a setting in which units in a network are assigned treatment status $T_i \in \{0,1\}$, as in \cite{aronow2017estimating}.  The network is a graph, $\mathcal{G}$, consisting of individuals (nodes) and connections (edges). For now, we consider the case in which treatment assignments are independent across nodes. 
 However, there are spillovers in treatment effects determined by the topology of the network where treatment status within one's network neighborhood may influence one's own outcome - for instance whether a friend is vaccinated affects a person's chance of being exposed to a disease.

Rather than being arbitrary, \cite{aronow2017estimating} consider an exposure function $f$ 
that takes on one of $K$ finite values; i.e.,  $f(T_i; T_{1:n},\mathcal{G}) \in \{d_1,\ldots, d_K\}$. An estimand of the average causal effect is of the form
\[
\tau(d_k,d_l) = \frac{1}{n}\sum_i y_i(d_k) - \frac{1}{n}\sum y_i(d_l)
\]
where $d_k$ and $d_l$ are the induced exposures under the treatment vectors. 
The Horvitz and Thompson estimator from \cite{horvitz1952generalization} provides:
\begin{align*}
    \hat{\tau}_{HT}(d_k,d_l) = \frac{1}{n}\sum_i \mathbf{1}\{ D_i = d_k \} \frac{y_i(d_k)}{\pi_i(d_k)} - \frac{1}{n}\sum \mathbf{1}\{ D_i = d_l \} \frac{y_i(d_l)}{\pi_i(d_l)}
\end{align*} where $\pi_i(d_k)$ is the probability that that node $i$ receives exposure $d_k$ over all treatments.

The challenge is that the treatment effects are not independent across subjects.  
Let $N_{i,:}$ denote a dummy vector of whether $j$ is in $i$'s neighborhood: let $N_{ij}=1$ when $G_{ij}=1$ with the convention $N_{ii}=0$. 
\cite{aronow2017estimating} consider an empirical study with $K = 4$ that are: (i) only $i$ is treated in their neighborhood ($d_1 = T_i \cdot 1\{T_{1:n}'N_{i ,:}=0\}$), (ii) at least one is treated in $i$'s neighborhood and $i$ is treated ($d_2 = T_i \cdot 1\{T_{1:n}'N_{i ,:}>0\}$), (iii) $i$ is not treated but some member of the neighborhood is ($d_3 = (1-T_i) \cdot 1\{T_{1:n}'N_{i ,:}>0\}$), and (iv) neither $i$ nor any neighbor is treated ($d_4 = (1-T_i) \cdot \prod_{j:\ N_{ij}=1}(1-T_j)$). We show that our result allows for a more generalized setting.

\subsubsection{Application of Theorem \ref{clt}}
To obtain consistency and asymptotic normality \cite{aronow2017estimating} assume a covariance restriction of local dependence (Condition 5) and apply \cite{chen2004normal} to prove the result.  Namely, their restriction is that there is a dependency graph $H$ (with entries $H_{ij}\in \{0,1\}$) with degree that is uniformly bounded by some integer $m$ independent of $n$. That is,  $\sum_{j} H_{ij} \leq m$ for every $i$. This setting is much more restrictive than our conditions, especially as there can exist indirect correlation in choices as effects propagate or diffuse through the graph. We can work with larger real exposure values, and in settings concentrating the mass of influence in a neighborhood while allowing from spillovers from everywhere. This is important to allow for in centrality-based diffusion models, SIR models, and financial flow networks, since the the spillovers in these settings are less restricted than the sparse dependency graph in their Condition 5.

Indeed, we can even allow the dependency graph to be a complete graph, as long as the correlations between the nodes in this dependency graph satisfy our Assumptions \ref{one}-\ref{three}. That is, we can handle cases where, for a given treatment assignment, each node has $n$ real exposure conditions for which the exposure conditions of the whole graph can be well approximated by simple functions, where small perturbations to any node in large regions of small correlations do not substantially perturb the outcomes in these regions (i.e., across affinity sets) while perturbations of the same size in any region of larger correlations (i.e., within affinity sets) can cause significant changes in the outcomes in that region. 
One can think of the ``shorter" monotonic regions in the simple function to be over affinity sets, and ``longer" monotonic regions to be across different affinity sets. One can think about monotonically non-decreasing functions, for instance, in epidemic spread settings where any increase in the ``treatment" cannot decrease the number of infected nodes. 

To take an example, let the true exposure of $i$ be given by $e_i(T_{1:n})$. Then consider a case where  $e_i(T_{1:n}^{+}) - e_i(T_{1:n}) \geq 0$, where $T_{1:n}^{+}$ indicates an increase in any element $j \in [n]$ from $T_{1:n}$. This is a structure that would happen naturally in a setting with diffusion. The potential outcome for $i$ given treatment assignment is assumed to be $y_i(e_i)$.

In practice, for parsimony and ease exposures are often binned. So consider  the problem where the $2^n$ possible exposures $e_i(T_{1:n})$ can be approximated by $K$ well-separated ``effective" exposures $\{d_1, d_2, ..., d_K\}$ where $|d_i - d_j| > \delta$ for any $i,j \in \{1, 2, ..., K \}$ and some $\delta > 0$, and  for any $r \in \{1,2,..., K \}$, $i,j \in \{1, 2, ..., 2^n\}$, we have, $e_i(T_{1:n}), e_j(T_{1:n}) \in d_r$ if and only if $| e_i(T_{1:n}) - e_j(T_{1:n})| < \delta$ and we have $y_i(e_i)$ smooth in its argument for every $i$. 

Then, following the above, the researcher's target estimand is the average causal effect switching between two exposure bins,
\[
\tau(d_k,d_l) = \frac{1}{n}\sum_i 1\{e_i(T_{1:n}) \in d_k\} y_i(e_i) - \frac{1}{n}\sum_i 1\{e_i(T_{1:n}) \in d_l\} y_i(e_i).
\]
The estimator for this estimand cannot directly be shown to be asymptotically normally distributed using the prior literature. It is ruled out by Condition 5 in \cite{aronow2017estimating} which uses \cite{chen2004normal}. However, it is straightforward to apply our result.

An example of this is a sub-critical diffusion process with randomly selected set of nodes $M_n$ being assigned some treatment, and every other node is subsequently infected with some probability. We provide two examples which speak to this in Subsections \ref{sec:sub-critical} and \ref{sec:sbm}.

\subsection{Covariance Estimation using Socio-economic Distances}
\subsubsection{Environment}
One application of mixing random fields is  to use them to develop covariance matrices for estimators (e.g., \cite{driscoll1998consistent,bester2011inference,barrios2012clustering,cressie2015statistics}). Here we consider the example of \cite{conley2002socio}, which builds on \cite{conley1999gmm}.  Essentially, their approach is to parameterize the characteristics (observable or unobservable) of units that drive correlation in shocks by the Euclidean metric, as we further describe below. This, however, rules out examples that are common in practice that include (discrete) characteristics with no intrinsic ordering driving degrees of correlation. For instance, correlational structures across race, ethnicity, caste, occupation, and so on, are not readily accommodated in the framework. For a concrete example,  correlations between ethnicities $e_i$ and $e_j$ for units $i,j$ that are parametrized by $p_{e_ie_j}$ in an unstructured manner are ruled out. Like this, many of these examples only admit partial orderings, if that. Yet these are important, practical considerations in applied work. The results in our Theorem \ref{clt} allows an intuitively nice treatment of such cases. Our discussion below also applies to combinations of temporal (and possibly cross-sectional) dependence as in \cite{driscoll1998consistent}.
Our conditions also provide consistent estimators for covariance matrices of moment conditions for parameters of interest in the GMM setting, under full-rank conditions of expected derivatives \cite{conley1999gmm}, since the author uses the CLT from \cite{bolthausen1982} under stationary random fields which is generalized above. 

 In \cite{conley1999gmm}, the model is that the population lives in a Euclidean space (taken to be $\mathbb{R}^2$ for the purposes of exposition), with each individual $i$ at location $s_i$. Each location has an associated random field $X_{s_i}.$ \cite{conley1999gmm} obtains the limiting distribution of parameter estimates $b$ of $\beta \in B$, where $B \subset \mathbb{R}^d$ is a compact subset and $\beta$ is the unique solution to $\E[g(X_{s_i}; \beta)]$ for a moment function $g$. \cite{conley1999gmm} lists sufficient conditions on the moment function to imply consistent estimation of the expected derivatives and having full rank:
\begin{enumerate}
    \item[a)] for all $b \in B$, the derivative $Dg(X_{s_i}; b)$  with respect to $b$ is measurable, continuous on $B$ for all $X \in \mathbb{R}^k$, and first-moment continuous.
    \item[b)] $\E[Dg(X_{s_i}; b)] < \infty$ and is of full-rank.
    \item[c)] $\sum_{s\in \mathbb{Z}^2} \cov (g(X_0; \beta), g(X_s;\beta))$ corresponding to sampled locations $X_s$ is a non-singular matrix.
\end{enumerate}

In addition to the sufficient conditions on the expected derivatives above, we list the remaining sufficient conditions on the random field $X_s$ itself used by \cite{conley1999gmm} to obtain the limiting distribution of parameter estimates through the GMM; 
and these are nested in the conditions from \cite{jenish2009central}, with the addition of bounded $2+\delta$-moment of $||g(X_s; \beta)||$:
\begin{enumerate}
    \item $\sum_{m=1}^\infty m \alpha_{k,l}(m) < \infty$ for $k + l \leq 4$
    \item $\alpha_{1, \infty}(m) = o(m^{-2})$
    \item for some $\delta >0$, $\E[||g(X_s; \beta)||]^{2 + \delta} < \infty$ and $\sum_{m=1}^\infty m (\alpha_{k,l}(m))^{\delta/(2 + \delta)} < \infty$.
\end{enumerate}

In \cite{conley2002socio}, the authors develop consistent covariance estimators, using these conditions, combining different distance metrics including physical distance as well as ethnicity (or occupation, for another example) distance in $L_2$ at an aggregate level (using census tracts data). In particular, the authors use indicator vectors to encode ethnicities (or occupations), and take the Euclidean distance from aggregated (at the census tract level) indicator vectors, as a measure of these ethnic/occupational distances. They use the Euclidean metric to write a ``race and ethnicity" distance between census tracts $i$ and $j$,
\begin{align*}
    D_{ij} = \sqrt{\sum_{k=1}^9 (e_{ik} - e_{jk})^2},
\end{align*} where the sum is taken over nine ethnicities/races, indexed by $k$, defined by the authors.

The use of indicator vectors and Euclidean distance results people of different race/ethnicity groups being in orthogonal groups (with a fixed addition of Euclidean distance $\sqrt{2}$ between any pairs of different race/ethnicity groups). To apply in practice, one would often need to allow for varying degrees of pairwise correlation for each pair of race/ethnicity groups. Additionally, even if the correlation induced by physical distance vanishes, it may be of interest to us to maintain correlation arising from interactions within and between ethnicity groups, where being of similar ethnic groups may induce nontrivial transfer of information between people despite being physically located large distances apart. It is not too difficult to see that the indicator vector formulation above does not allow for this, since in the case of a pair of distinct ethnicities with high correlation, the formulation in \cite{conley2002socio} would require a correlation of zero.

\subsubsection{Application of Theorem \ref{clt}}

Consider random variables $Z_i$ and $Z_j$, whose correlation we can decompose into components of physical distance and racial/ethnic distance (just as in \cite{conley2002socio}). It is direct to see how our work above takes care of the physical distance component, and so we turn to the remaining distance component. For this, for instance, one could consider the pairwise interaction probabilities, $p_{e_i,e_j}$ characterizing the correlation between ethnicity $e_i$ of $i$, and ethnicity $e_j$ of $j$. Our affinity set structure then allows us to incorporate this correlation structure. That is, one can construct an affinity set $\A_i^n = \{j: \rho(i,j) \leq K^i(\epsilon/2), \text{ or } p_{e_i,e_j} \geq 1- \epsilon/2 \}$ with $K^i$ defined just as in Subsection \ref{rf}. Following our previous section, our generalization holds.

Attempts to (non-parametrically) estimate covariance in the cross-section often leverage a time or distance structure. For example, \cite{driscoll1998consistent} assume a mixing condition on a random field such that the correlation between shocks $\epsilon_{it}$ and $\epsilon_{j,t-s}$ tends to zero as $s \rightarrow \infty$.  This allows for reasonably agnostic cross sectional correlational structures but requires it to be temporally invariant and studies $T^{1/2}$ asymptotics. 
Although such an assumption applies in certain contexts, there are many socio-economic contexts in which it does do not apply and yet our theorem can be applied. We provide two such examples.

For instance, in simple models of migration with migration cost, there is often persistence in how shocks to incentives to migrate in some areas affect populations in other areas. Nonetheless, there are very particular correlation patterns because, as an example, ethnic groups migrate to specific places based on existing populations, and so affinity sets are driven by the places to which a given ethnic group might consider moving and our central limit theorem can then be applied, provided our conditions on affinity sets apply.
 
Another example comes from social interaction. Individuals interact with others in small groups that experience correlated shocks which correlate grouped individuals' behaviors and beliefs.  Each group involves only a tiny portion of the population, and any given person interacts in a series of groups over time.   Thus individuals' behaviors or beliefs are correlated with others with whom they have interacted; but without any natural temporal or spatial structure.  Each person has an affinity set composed of the groups (classes, teams, and so on) that they have been part of. People may also have their own idiosyncratic shocks to behaviors and beliefs. In this example again, our central limit theorem applies despite the lack of any spatial or temporal structure.  One does not need to know the affinity sets, just to know that each person's affinity group is appropriately small relative to the population.

\subsection{Sub-Critical Diffusion Models} 
\label{sec:sub-critical}
\subsubsection{Environment}  
A finite-time SIR diffusion process occurs on a sequence of unweighted and undirected graphs $G_n$. A first-infected set, or seed set $M_n$, of size $m_n$, of random nodes with treatment indicated $W_i \in \{0,1\}$, are seeded (set to have $W_i=1$) at $t=0$ and in period $t=1$ each infects each of its network neighbors, $\{j:\ G_{ij,n}=1\}$ i.i.d. with probability $q_n$. The seeds then are no longer active in infecting others. In period $t=2$ each of the nodes infected at period $t=1$ infects each of its network neighbors who were never previously infected i.i.d. with probability $q_n$. The process continues for $T_n$ periods. Let $X_i^n \in \{0,1\}$ be a binary indicator of whether $i$ was ever infected throughout the process.

Assume that the sequence of SIR models under study, $(G_n,q_n,T_n,W_n)$
have $m_n \rightarrow \infty$ (with $\alpha_n:= m_n/n = o(1)$), $q_n \rightarrow 0$, and $T_n \rightarrow \infty$ (with $T_n \geq \text{diam}(G_n)$ at each $n$), and are such that the process is sub-critical. Since the number of periods is at least as large as the diameter, it guarantees that for a connected $G_n$, $\cov(X_i^n,X_j^n) > 0$ for each $i,j$.  

The statistician may be interested in a number of quantities. For instance, the unknown parameter $q_n$ may be of interest. Suppose 
 $\E[\Psi_i(X_i;q_n,W_{1:n})]= 0$ is a (scalar) moment condition satisfied only at the true parameter $q_n$ given known seeding $W_{1:n}$. The $Z$-estimator (or GMM) derives from the empirical analog, setting 
\(
\sum_{i} \Psi_i(X_i; \hat q,W_{1:n}) = 0.
\) 

By a standard expansion argument
\begin{align*}
(\hat q - q_n) = -\left\{\sum_i \nabla_q \Psi_i(X_i; \tilde q, W_{1:n})\right\}^{-1} \times \sum_{i} \Psi_i(X_i; q_n,W_{1:n}) + o_p(n^{-1/2}).
\end{align*}
To study the asymptotic normality of the estimator, we need to study
\[
\frac{1}{\sqrt{\var\left(\sum_{i} \Psi_i \right)}}\sum_{i} \Psi_i(X_i; q_n,W_{1:n}),
\]
which involves developing affinity sets for each $\Psi_i$.
For simplicity  we consider the case where the estimator may directly work with $X_{1:n}^n$. Letting $Z_i = X_i^n - \E(X_i^n)$ be the de-meaned outcome and we want to show that
 \begin{align*}
\frac{1}{\sqrt{\sum_i \sum_{j \in \A_i^n} \cov(Z_i^n,Z_j^n)}}\sum_i Z_i^n \wkto \mathcal{N}(0,1).
 \end{align*}
 
Under sub-criticality, a vanishing share of nodes are infected from a single seed. Without sub-criticality, most of the graph can have a nontrivial probability of being infected and accurate inference cannot be made with a single network.
Let us define
\begin{align*}
  \B_{j}^n :=  \{ i: \Pr(X_i^n = 1 \mid j \in M_n, m_n = 1) > \epsilon_n \}.
\end{align*}
Then $ \B_j^n  $ is the set of nodes for which, if $j$ is the only seed, the probability of being infected in the process is at least $\epsilon_n$. As noted above, in a sub-critical process $\left\vert \B_{j}^n \right\vert = o(n)$ for every $j$, not necessarily uniformly in $j$. For simplicity assume that there is a sequence $\epsilon_n \rightarrow 0$ such that this holds uniformly (otherwise, one can simply consider sums). Let $\beta_n := \sup_j \left\vert \B_{j}^n \right\vert/n$ which tends to zero, such that $\alpha_n \beta_n = o(n)$.

Next, we assume that the rate at which infections happen within the affinity set is higher than outside of it, and the share of seeds is sufficiently high and affinity sets are large enough to lead to many small infection outbursts but not so large as to infect the whole network. That is,   there exists some $\B_{j}^n {}' \subset \B_{j}^n$ such that $|\B_{j}^n{}'| = \Theta(|\B_{j}^n|)$ and $\Pr(X_i = 1 | j \in M_n, m_n = 1) \geq \gamma_n$ with $\gamma_n /\epsilon_n \rightarrow \infty$, such that $\alpha_n^3 = O(\gamma_n)$, and $\beta_n = O(\gamma_n)$.   This would apply to, for instance, targeted advertisements or promotions that lead to local spread of information about a product, but that does not go viral.
None of the prior examples such as random fields and dependency graphs cover this case since since all $X_i^n$ are correlated. We now show that Theorem \ref{clt} applies to this case.
\subsubsection{Application of Theorem \ref{clt}}
Let us define the affinity sets $\A_i^n = \B_{i}^n$. Next, consider a random seed $k$. Let $\mathcal{E}_{i,j,k} := \{a \notin \B_{b}^n:  \  a,b \in  \{i,j,k\},\ a\neq b\} $ denote the event that none of the nodes are in each others' affinity sets. It is clear that $\Pr(\mathcal{E}_{i,j,k}) \rightarrow 1$, since   $\left\vert \B_{a}^n\right\vert = o(n)$ for $a \in \{i,j,k\}$ and that seeds are uniformly randomly chosen.

If we look at an affinity set, it is sufficient to just look at the variance components and check it is a higher order of magnitude
\[
\sum_{i} \var(Z_i^n) = n \times (1-(1-\beta_n)^{2m_n}) \gamma_n^2 = O(n^2 \times \alpha_n \beta_n \times \gamma_n^2).
\]
Now to check Assumption \ref{one}, we compute:
\begin{align*}
    \sum_{i;j,k \in \A_i^n }\E[|Z_i| Z_j Z_k] &= \sum_{i;j,k \in \A_i^n, j \notin \A_k^n \text{ or } k \notin \A_j^n }\E[|Z_i| Z_j Z_k] + \sum_{i;j,k \in \A_i^n, \A_j^n, \A_k^n} \E[|Z_i| Z_j Z_k] \\
    &\approx n^3 \beta_n \epsilon_n^2 + n \alpha_n \beta_n \epsilon_n.
\end{align*}
Thus, we have 
\[ \frac{n^6 \alpha_n^3 \beta_n^3 \gamma_n^6}{n^6 \beta_n^3 \epsilon_n^6} = \frac{\alpha_n^3  \gamma_n^6}{\epsilon_n^6} \to \infty, \] which is satisfied.
The probability that no seed is in any other affinity set
\[
\Pr(\cap_{k \in M_n}\mathcal{E}_{i,j,k}) = (1-\beta_n)^{2m_n} \approx 1- 2m_n \beta_n = 1-2n \times \alpha_n \beta_n.
\]
This puts an intuitive restriction on the number of seeds and percolation size as a function of $n$. 
Next, we verify Assumption \ref{two}. We have,
\begin{align*}
    \sum_{i; j,k\in\A_i^n, r\in\A_k^n} \cov(Z_i^n Z_j^n, Z_k^n Z_r^n) &=  O( \sum_{i; j,k\in\A_i^n; r\in A_k^n, \notin A_i^n} \cov(Z_i^nZ_j^nZ_k^nZ_r^n) ) \\
    &= O(n^2 \times (n-1)(1-\beta_n)\beta_n  \epsilon_n^4) 
    =O( n^3 \beta_n \epsilon_n^4).
\end{align*}

Therefore, 
${(n^4 \alpha_n^2 \beta_n^2 \gamma_n^4)}/{(n^3 \beta_n \epsilon_n^4)} = n \alpha_n \beta_n (\frac{\gamma_n}{\epsilon_n})^4 \to \infty$,   
is satisfied.
We next verify Assumption \ref{three}. Given the event $\mathcal{E}_{i,j,k}$, we can bound the conditional covariance,
 $\cov(Z_i^n,Z_j^n \mid \mathcal{E}_{i,j,k}) = O(\epsilon_n^2)$,
 by bounding the probabilities of two contagions. So then
\begin{align*}
    \sum_{i,j: \ j \notin \A_i^n} \sum_{k \in M_n}\cov(Z_i^n, Z_j^n \mid \mathcal{E}_{i,j,k})  \Pr(\mathcal{E}_{i,j,k}) 
    &=  C \sum_{i,j: \ j \notin \A_i^n} (1-2n \alpha_n \beta_n) \cdot \cov(Z_i^n, Z_j^n \mid \mathcal{E}_{i,j,k}) \\
    &\approx (1-\alpha_n)\cdot n \times  ((1-\alpha_n)\cdot n -1) (1-\beta_n) \times (1-2 n \alpha_n \beta_n) \cdot \epsilon_n^2
\end{align*} for some constant $C>0$ fixed in $n$.
Keeping orders we have
\[
\sum_{i,j: \ i \notin \A_j^n} \sum_{k \in M_n}\cov(Z_i^n, Z_j^n \mid \mathcal{E}_{i,j,k})  \Pr(\mathcal{E}_{i,j,k})  = O((n \epsilon_n)^2).
\]
Since $ \sum_{i,j: \ i \notin \A_j^n} \sum_{k \in M_n} \cov(\E[Z_i^n \mid  \mathcal{E}_{i,j,k} ], \E[ Z_j^n \mid  \mathcal{E}_{i,j,k}])  = 0$, we have, $\sum_{i,j: \ i \notin \A_j^n} \sum_{k \in M_n}\cov(Z_i^n, Z_j^n )   = O((n \epsilon_n)^2)$ and 
${(n^2 \gamma_n^2 \alpha_n \beta_n})/({n^2 \epsilon_n^2)} = ({\gamma_n^2 \alpha_n \beta_n})/({\epsilon_n^2} )\rightarrow \infty,
$ is also satisfied.

\subsection{Diffusion in Stochastic Block Models} \label{sec:sbm}

\subsubsection{Environment} 

A SIR diffusion process occurs on a sequence of unweighted and undirected networks,
as in the previous example, except that the network has a block structure as generated by a standard stochastic block model (\cite{holland1983stochastic}).
The $n$ nodes are partitioned into $k_n$ blocks, where block sizes are equal, or within one of each other.   
With probability $p^{in}_n\in (0,1)$ links are formed inside blocks,  and with probability $p^{ac}_n\in (0,1)$ they are formed across blocks, independently.  
Let $q_n$ be the probability of infection, as in the last example.
Inside link probabilities are large enough for percolation within blocks:    $p^{in}_n \frac{n}{k_n} q_n>> \log\left(\frac{n}{k_n}\right)$.
Across link probabilities are small enough for vanishing probabilities of contagion across blocks, even if all other blocks are infected: $p^{ac}_n  n^2 q_n <<1$.
The infections are seeded with $k_n/2$ seeds.   With probability going to 1, all nodes in the blocks with the seeds will be infected and no others.  
There is a correlation going to 1 of infection status of nodes within the blocks,  which are the affinity sets; and there is correlation of infection status going to 0 across blocks, but it is always positive.
If $k_n$ is bounded, then a central limit theorem fails.   If $k_n$ grows without bound 
(while allowing $n/k_n\rightarrow \infty$ so that blocks are large), 
then the central limit theorem holds.   

\subsubsection{Application of Theorem \ref{clt}}

Given the previous examples, we simply sketch the application.
The affinity set  of node $i$, $\A_i^n$, is the block in which it resides.  
Letting $\sigma_n^2$ denote $\var(Z_i)$,  it follows that $\Omega_n\approx n\frac{n}{k_n} \sigma_n^2$.
Then the first assumption is satisfied noting that if $k_n\rightarrow \infty$,
then
\[
\sum_{i;j,k \in \A_i^n }\E[|Z_i| Z_j Z_k]=  O(  \sum_{i;j,k \in \A_i^n }\E[|Z_i|^3] ) =   O\left( n \left(\frac{n}{k_n}\right)^2 \E[|Z_i|^3]\right)= o(\Omega_n^{3/2} ).
\]
Verification of the second assumption comes from noting that if $k_n\rightarrow \infty$ then
\begin{eqnarray*}
    \sum_{i; j,k\in\A_i^n, r\in\A_k^n} \cov(Z_i^n Z_j^n, Z_k^n Z_r^n) &=&  O\left( \sum_{i; j,k,r\in\A_i^n} \cov(Z_i^n Z_j^n, Z_k^n Z_r^n) \right) \\
    &=& O\left(n \left(\frac{n}{k_n}\right)^3 \E[ Z_i^4]\right) =o( \Omega_n^{2}).
\end{eqnarray*}
Next we check the third assumption. Let $\varepsilon_n=\cov(Z_i^n,Z_j^n)$ for $i,j$ in different blocks (ignoring the approximation due to the fact that blocks may be of slightly different sizes).  Note that $\varepsilon_n=\cov(Z_i^n,Z_j^n)$ is on the order of contagion across blocks, which is $p^{ac}_n q_n ({n^2}/{k_n^2}) = o({1}/{k_n^2})$. Both blocks could also be infected by some other nodes, which happens of order at most $n^2 (p^{ac} q_n ({n}/{k_n}))^2$, which is also of order $o({1}/{k_n^2})$.    If $k_n$ grows without bound
\[
\sum_{i} \E \left(| \E[Z_{i}  \mathbf{Z}_{-i} \vert  \mathbf{Z}_{-i} ]| \right) = O\left( n^2\frac{k_n-1}{k_n} \varepsilon_n\right)
	= o\left( \frac{n^2}{k_n} \sigma^2_n \right).
\]

In this example, not only do the assumptions fail if $k_n$ is bounded, but also 
the conclusion of the theorem fails to hold as well; so the conditions are tight in that sense.   

\subsection{Spatial Process with Irregular Observations and Mat\'{e}rn Covariance}
\subsubsection{Environment} Finally, we turn to an example of neural network models for geospatial data. Specifically, we look at the environment of \cite{zhan2023neural}. The authors propose a neural network generalized least squares process (NN-GLS) with the dependency in the residuals modeled by a Mat\'{e}rn covariance function, described below. Their paper is the first to demonstrate consistency for the NN-GLS estimator in this setting. 

Consider a spatial process model,
$Y_i=f_0(X_i)+\varepsilon(s_i)$, where $X_i \in \mathbb{R}^k$ is a vector of characteristics and the residuals correspond to  observations at locations $s_1,..,s_n$ in $\mathbb{R}^2$.  Let $f_0(\cdot)$ be a continuous function and define $\varepsilon(s_i)$ as a Gaussian Process with covariance function $\Sigma(s_i,s_j)=C(s_i,s_j)+\tau^2\delta(s_i=s_j)$ for some $\tau^2>0$ and $\delta$ is the indicator function. Here $C(s_i, s_j) = C(||s_i-s_j||_2) = C(||h||_2)$, where $$C(||h||_2) = \sigma^2 \frac{2^{1-\nu}(\sqrt{2}\phi||h||_2)^\nu}{\Gamma(\nu)}\mathcal{K}_\nu(\sqrt{2}\phi||h||_2)$$ is the Matérn covariance function, with modified Bessel function  of the second kind $\mathcal{K}_\nu(\cdot)$. We consider the setting in \cite{zhan2023neural} (Proposition 1) where $C \left (||h||_2 \right ) = o\left({||h||_2}^{-(2+ \kappa)}\right)$ for some $\kappa>0$.

The NN-GLS fits a system of multi-layered perceptrons via the $L_2$ loss function and the authors prove consistency under some assumptions including, in particular, restrictions on the spectral radius of a sparse approximation of the covariance function.  That is covered by an assumption of minimum distance $h_0 > 0$ separation of locations $s_i, s_j$ above, where $i \neq j$. Previous work characterizes the asymptotic properties, including asymptotic normality of the neural network estimators, in the case of independent and identically distributed shocks \citep{shen2019asymptotic}. \cite{zhan2023neural} extend this result by modeling dependency using the Mat\'{e}rn covariance function.

\subsubsection{Application of Theorem~\ref{clt}} 
We create affinity sets using the same restrictions as presented in~\cite{zhan2023neural}. Reflecting the duality of spatial distance and dependence, we construct affinity sets such that the maximal separation in the affinity set has implications for the maximum covariance between random variables.  Specifically, take the affinity sets to be defined as $\A_{i}^n := \{j: ||s_i - s_j||_2 <  K(\epsilon_n)\}$ where $|\cov(Z_{i,n},Z_{j,n})| \leq \epsilon_n$ for  $||s_i - s_j||_2 >  K(\epsilon_n)$. Using ~\cite{zhan2023neural}'s restriction on the amount of dependence associated with the distance in space, namely $C \left (||h||_2 \right ) = o\left({||h||_2}^{-(2+ \kappa)}\right)$ for some $\kappa>0$, we can solve for the appropriate $K(\epsilon_n)$.  Specifically, if we know that the covariance in~\cite{zhan2023neural} is asymptotically bounded by ${||h||_2}^{-(2+ \kappa)}$, we can set this equal to $\epsilon_n$ and solve for the appropriate distance.  After the resulting algebra we take $K(\epsilon_n)=1/(\epsilon_n^{2+\kappa})$.  

Until now, we have defined distances that give affinity sets that contain the bulk of the dependence.  We also need to ensure that, under the setup in~\cite{zhan2023neural}, these sets are not too large that they violate our assumptions.  To do this, we 
take $\epsilon_n = \omega( \frac{1}{h_0^2 n^{\gamma}})$ for $0 < \gamma < 1$ and $h_0$ as the minimum separation distance, defined above. 
Using a packing number calculation, we see that while this allows $K(\epsilon_n)$ to grow with $n$, it grows more slowly than $n$. Specifically, we have, 
\begin{align*}
    \Big(\frac{K(\epsilon_n)}{h_0}\Big)^2 = \Big(\frac{(1/\epsilon_n)^{-(2 + \kappa)}}{h_0}\Big)^2 < \Big(\frac{1}{\epsilon_n h_0^2} \Big) = o(n).
\end{align*}

This logic generalizes to dimensions $d \geq 1$, taking $\epsilon_n = \omega(\frac{1}{h_0^dn^\gamma})$ for $0 < \gamma < 1$.
Using this construction and assuming bounded third and fourth moments, we check that our Assumptions \ref{one}-\ref{three} apply.

Letting $K:= K(\epsilon)$, we show Assumption \ref{one} holds since 
\begin{eqnarray*}
    \sum_{i; j, k \in \A_i^n} \E[|Z_{i,n}| Z_{j,n} Z_{k,n}]  &\leq&\sum_{i; j, k \in \A_i^n} \E[|Z_{i,n}|| Z_{j,n}| Z_{k,n}|] \\ &\leq&
    \sum_{i; j, k \in \A_i^n} \left(\frac{1}{3}\E[|Z_{i,n}|^3]+\frac{1}{3}\E[|Z_{j,n}|^3]+\frac{1}{3}\E[|Z_{k,n}|^3]\right)\\
    &=&
    O\left(\sum_{i; j, k \in \mathcal{A}_i^n} \E[|Z_{i,n}|^3]\right)\\
    &=&O(K^2 \sum_{i} \E[|Z_{i,n}|^3])
    =O(K^2n)
    =o(n^{3/2}K^{3/2})
    =o\left(\Omega_n^{3/2}\right).
\end{eqnarray*} 
 The second inequality holds by the algorithm-geometric mean inequality.  The remaining argument relies on rearranging the summations and using the growth rate of $K$, i.e. $K = o(n)$ in the fourth equality, as defined above based on the conditions required by~\cite{zhan2023neural}.The last equality follows since $\Omega_n^{3/2} = (\sum_{i} \sum_{j \in \A_i^n} \E[Z_iZ_j])^{3/2}.$

We check that Assumption \ref{two} is satisfied using similar arguments and relying on an assumption of finite fourth moment:
\begin{eqnarray*}
\sum_{i,j;k\in \A_i^n,l\in \A_j^n} \cov(Z_{i,n}Z_{k,n},Z_{j,n}Z_{l,n}) &=& O( \sum_{\mathclap{\substack{i,j: |s_i - s_j| \leq K,\\
k: |s_k - s_i| \leq K,
l: |s_l - s_i| \leq 2K}}} \var(Z_{i,n}^2)) = O(n \cdot K^3 \cdot \E(Z_{i,n}^4)) = o(n^2 K^2) = o(\Omega_n^2).
\end{eqnarray*} The first equality comes from the construction of affinity sets such that the covariance terms within the affinity sets dominate those with outside the affinity sets (additional details in Assumption \ref{three}). The remaining equalities use similar arguments to the Assumption above.

Assumption \ref{three} follows from taking $\epsilon_n = \omega({1}/({h_0^dn^\gamma}))$ where $\gamma =  1 - \beta$ for arbitrarily small $\beta > 0.$ Indeed, taking $\epsilon_n$ as such, we have, ${n}/({K^{1 + 1/(2+\kappa)}}) = o(1)$, and thus,  

\begin{align*}
      \sum_{i} \E ( |\mathbf{Z}_{-i,n} \E (Z_{i,n}  | \mathbf{Z}_{-i,n})|  ) 
    &= O \left( \sum_{i} \frac{\epsilon_n}{2} \E ( |\mathbf{Z}_{-i,n}  Z_{i,n}| ) \right) \\
    &\leq O \left( \sum_{i} \sum_{j \notin \mathcal{A}_i}  \epsilon_n \E ( |Z_{j,n}  Z_{i,n}|) \right) \\
    &= O \left( \sum_{i} \sum_{j \notin \mathcal{A}_i}  \epsilon_n \int_0^1 Q_Z(u)^2 du \right) \\
    &= O(n(n-K) \epsilon_n)
    = O(n(n-K) K^{-1/(2 + \kappa)}) = o(nK)   
    = o\left(\Omega_n\right),
\end{align*} where the first line above is obtained by observing that for jointly Gaussian random variables $X_1, X_2$, we can write $\mathbb{E}[X_2 | X_1] = \mu_2 + \frac{\cov(X_1, X_2)}{\var(X_1)}(X_1 -\mu_1)$, where $0 < \var(X_1) < \infty$ due to the eigenvalues of the covariance matrix being uniformly bounded in $n$.

\section{Discussion}
\label{sec:discussion}

We have provided an organizing principle for modeling dependency and obtaining a central limit theorem: affinity sets. It allows for non-zero correlation across all random vectors in the triangular array and places focus on correlations within and across sets. These conditions are intuitive and we illustrate their use through some practical applications for applied research. In some cases, as in several of our applied examples, our result is needed as previous conditions do not apply. 

We now reflect on settings that our theorem does not cover. For example, 
the martingale central limit theorem  
(e.g., \cite{billingsley1961lindeberg,ibragimov1963central,hall2014martingale} among others)
is not covered by our theorem, without modification.  It admits nontrivial unconditional correlation between all variables, but relies on other structural properties to deduce the result.
In fact, proofs of the martingale central limit theorem did not appeal to Stein's method, until \cite{rollin2018quantitative}. By combining Stein and Lindeberg methods, \cite{rollin2018quantitative} develops a shorter proof but did not find a direct proof using the Stein technique alone.
Some biased processes that do not fall under the martingale umbrella can still generate a central limit theorem if they satisfy the covariance structure that we have provided. 

\bibliographystyle{abbrvnat}
\bibliography{metrics}

\pagebreak
\begin{appendix}
~\\
\centerline{\Large Appendix}
\section*{Proof of Central Limit Theorem \ref{clt} and Corollary \ref{cor:singleton} }\label{sec:normality}
Recall that  $\Omega_n$ is a  $p \times p$ matrix with entries
      \begin{align*}
         \Omega_{n,dd'} := \sum_{i = 1}^n \sum_{(j,d')\in \A^n_{(i,d)} } \cov \left(Z_{i,d}, Z_{j,d'}\right).
     \end{align*}

We start with the case $p = 1$, reproducing elements of the proof to Theorem 2 in \cite{chandrasekharj2018}. Let $\Omega_{n} := \sum_{i = 1}^n \sum_{j\in \A^n_{i} } \cov \left(Z_{i}, Z_{j}\right).$

The proof uses Stein's lemma from \cite{stein1986approximate}. 
\begin{lem}[\cite{stein1986approximate,ross2011fundamentals}]\label{lem:Stein} \ \\
	If $Y$ is a random variable and $Z$ has the standard normal distribution, then
	\[
	d_W(Y,Z) \leq  \sup_{\{f :  ||f||, ||f''||\leq  2, ||f'||\leq \sqrt{2/\pi} \}
	}
	\left|
	\E [f'(Y) - Y f(Y)]\right|.
	\]
	Further
	\(
	d_K(Y,Z)\leq (2/\pi)^{1/4} (d_W(Y,Z))^{1/2}.
	\)
\end{lem}

By this lemma, if we show that a normalized sum of random variables satisfies
\[
\sup_{\{f :  ||f||, ||f''||\leq  2, ||f'||\leq \sqrt{2/\pi} \}}
\left|
\E [f'(\overline{S}^n) - \overline{S}^n f(\overline{S}^n)]\right| \rightarrow 0,
\]
then $d_W(\barS^n,Z)\rightarrow 0$, and so it must be asymptotically normally distributed.

The following lemmas are useful in the proof.

\begin{lem}[\cite{chandrasekharj2018}, Lemma B.2] \label{lem:h_max} \ \\
A solution to $\max_h \E[Zh(Y)]   \mbox{ s.t. }  |h|\leq 1$ (where $h$ is measurable) is $h(Y) = \sign( \E[Z|Y] ),$ where we break ties, setting $\sign( \E[Z|Y] )=1$ when $ \E[Z|Y] =0$.
\end{lem}

\

\begin{lem}[\cite{chandrasekharj2018}, Lemma B.3] \label{lem:XY_h} \ \\
\( \E[XYh(Y)] \)  when \( h(\cdot) \) is measurable and bounded by \( \sqrt{\frac{2}{\pi}} \)  satisfies
\[ \E[XYh(Y)] \leq   \sqrt{\frac{2}{\pi}} \E\left[XY \cdot \sign(\E[X|Y] Y)  \right].\]
\end{lem}

\
\begin{lem}[Cram\'er-Wold Device, \cite{cramer1936some}] \label{lem:cramer_wold}
The sequence $\{X_n\}_n$ of random vectors of dimension $p \in \mathbb{N}$ weakly converges to the random vector $X \in \mathbb{R}^p$, as $n \to \infty$, if and only if for any $c \in \mathbb{R}^p$, 
\begin{align*}
    c^\intercal X_n \wkto c^\intercal X
\end{align*} as $n \to \infty.$
\end{lem}

\

\begin{proof}[{\bf Proof of Theorem \ref{clt}}] The $p=1$ case  is Theorem 2 in \cite{chandrasekharj2018}. So we reproduce a sketch to provide intuition.  
	By Lemma \ref{lem:Stein}, it is sufficient to show that the appropriate sequence of random variables $\overline{S}^n$ satisfies
	\[
	\sup_{\{f :  ||f||, ||f''||\leq  2, ||f'||\leq \sqrt{2/\pi} \}
	}
	\left|
	\E [f'(\overline{S}^n) - \overline{S}^n f(\overline{S}^n)]\right| \rightarrow 0.
	\]
	
Let
	\[
	S_i := \sum_{j \notin \A_i^n}Z_j  \mbox{ and } \bar{S}_i:=S_i/\Omega_n^{1/2}.
	\]
 Let $\overline{S}^{n}=S^{n}/\Omega_{n}^{1/2}.$

We consider
\begin{align}
    | \E [f'(\overline{S}^n) - \overline{S}^n f(\overline{S}^n)]| \label{lhs_stein}
\end{align}
for all $f$ such that $||f||, ||f''||\leq  2, ||f'||\leq \sqrt{2/\pi}$.   
Observe that
\begin{align}
\E\left[\bar{S}f\left(\bar{S}\right)\right] & =  \E\left[\frac{1}{\Omega_n^{1/2}}\sum_{i}^nZ_{i}\cdot f\left(\bar{S}\right)\right]
= \E\left[\frac{1}{\Omega_n^{1/2}}\sum_{i}^nZ_{i}\left(f\left(\bar{S}\right)-f\left(\bar{S}_{i}\right)\right)\right]
+ \E\left[\frac{1}{\Omega_n^{1/2}}\sum_{i}^nZ_{i}\cdot f\left(\bar{S}_{i}\right)\right]. \label{lhs_stein_second_term}
\end{align}

 We first consider the second term above. By a first-order Taylor approximation of $f\left( \overline{S} \right)$ about $f(0)$, and observing $\E[Z_i]=0$ and the triangle inequality, we get an upper bound:
 \begin{align*}
 \left|\E\left[ \Omega_n^{-1/2}\sum_i^n Z_i \cdot f\left( \overline{S}_i\right) \right] \right| &\leq
 \left|\E\left[ \Omega_n^{-1/2}\sum_i^n Z_i \cdot f\left(  0 \right) \right] \right| + 
 \left|\E\left[ \Omega_n^{-1/2}\sum_i^n Z_i \cdot (\overline{S}_i)f'\left( \tilde{S}_i\right) \right] \right|
 \end{align*} where $\tilde{S}_i$ is an intermediate term between 0 and $\overline{S}_i$.
 The first term is zero since $\E[Z_i]=0$. We upper bound the second term by applying Lemma \ref{lem:XY_h}. 
 \begin{align*}
      \left|\E\left[ \Omega_n^{-1/2}\sum_i^n Z_i \cdot \left( \Omega_n^{-1/2}\sum_{j \notin \A_i^n} Z_j \right) f'\left( \tilde{S}_i\right) \right] \right| &\leq \left|\Omega_n^{-1}\sqrt{2/\pi } \cdot \E\left[\sum_i Z_i \mathbf{Z}_{-i} \cdot \mathrm{sign}(\E[Z_i \mid \mathbf{Z}_{-i}] \mathbf{Z}_{-i}) \right] \right| \\
      &= \left| \Omega_n^{-1}\sqrt{2/\pi } \cdot  \E\left[\sum_i \mid \E\left[Z_i \mathbf{Z}_{-i} \mid \mathbf{Z}_{-i} \right] \mid \right] \right|\\
    &= \left|\Omega_n^{-1}\sqrt{2/\pi } \cdot  \E\left[\sum_i \mid \mathbf{Z}_{-i} \E\left[Z_i  \mid \mathbf{Z}_{-i} \right] \mid \right] \right|
 \end{align*}
  By Assumption \ref{three}, we have that the upper bound  is $o(1)$. This kind of bound is generated by the fact that, in principle, any two random variables can be correlated for a given $n$.

  Therefore, \eqref{lhs_stein_second_term} is simply 
  \begin{align*}
      \E\left[\frac{1}{\Omega_n^{1/2}}\sum_{i}^nZ_{i}\left(f\left(\bar{S}\right)-f\left(\bar{S}_{i}\right)\right)\right]
+ o(1).
  \end{align*}
	
Now, by plugging this expression into \eqref{lhs_stein} we have an upper bound by using the triangle inequality (we follow a reasoning similar to that in \cite*{ ross2011fundamentals}). Follow this by a second-order Taylor series approximation and a bound on the derivatives of $f$ and the use of the Cauchy-Schwarz inequality, together with Assumptions \ref{one} and \ref{two}. In both of these pieces, rather than relying on conditional independence and applying the arithmetic-geometric mean inequality to write conditions in terms of moment restrictions, which cannot be used with the general dependency structure, we collect covariance terms.

Therefore, we have shown the convergence $\overline{S}^n \wkto N(0,1)$ in distribution in each dimension. Now, we consider the multidimensional setting, and let $Y = (Y_1, Y_2, ..., Y_p)$ be a mean-zero normally distributed random vector with covariance the $p$-dimensional identity matrix. By the Cram\'er-Wold device (Lemma \ref{lem:cramer_wold}), it is sufficient to show that 
\begin{align}
    \sum_{u=1}^p c_u \sum_{i=1}^n \sum_{k=1}^p Z_{ik}{({\Omega_n}^{-1/2})_{ku}} \wkto   \sum_{u=1}^p c_u \sum_{i=1}^n Y_{iu} \label{cramer_condition}
\end{align}
for all $c \in \mathbb{R}^p$. 

But, from the proof above, we see that for each $u \in [p]$, $$\sum_{i=1}^n \sum_{k=1}^p Z_{ik}{({\Omega_n}^{-1/2})_{ku}} \wkto \sum_{i=1}^n Y_{iu}.$$ It immediately follows that \ref{cramer_condition} is satisfied, so we have shown $(\Omega_n)^{-1/2}S^n \wkto \mathcal{N}(0,I_{p \times p})$.\end{proof}

\

\begin{proof}[{\bf Proof sketch of Corollary \ref{cor:singleton}}] We refer the reader to the proof to Corollary 2 in \cite{chandrasekharj2018}. Applying the Cram\'er-Wold device just as above gives us the result.\end{proof}
\end{appendix}

\end{document}